\documentclass[aps, prb, superscriptaddress, reprint, amsmath, amssymb, showpacs]{revtex4-1}
\usepackage{graphicx}
\usepackage{dcolumn}
\usepackage{bm}
\usepackage{amsmath}

\begin{document}

\title{Nodal and nodeless gap in proximity induced superconductivity:
  application to monolayer CuO$_2$ on BSCCO substrate}
\author{Yimeng Wang}
\affiliation{Institute for Quantum Science and Engineering and Department of Physics, Southern University of Science and Technology, Shenzhen, China}
\author{Zhen-Hua Wang}
\email[]{wangzh@sustc.edu.cn}
\affiliation{Institute for Quantum Science and Engineering and Department of Physics, Southern University of Science and Technology, Shenzhen, China}
\affiliation{Beijing Computational Science Research Center, Beijing 100193,
  China}
\author{Wei-Qiang Chen}
\email[]{chenwq@sustc.edu.cn}
\affiliation{Institute for Quantum Science and Engineering and Department of Physics, Southern University of Science and Technology, Shenzhen, China}
\begin{abstract}
  We present a detailed analysis on the hopping between monolayer CuO$_2$ and
  bulk CuO$_2$ plane in the Bi$_2$Sr$_2$CaCu$_2$O$_{8+\delta}$ substrate. With a
  two-band model, we demonstrate that the nodeless gap can only exist when the
  hole concentration in monolayer CuO$_2$ plane is very large. We argue that the
  possible phase separation may play important role in the recent experimental
  observation of nodeless gap.
\end{abstract}

\maketitle
\section{INTRODUCTION}
The high temperature cuprate superconductor is one of the most important fields
in past thirty years\cite{muller_bednorz, RVB, Zhang_rice, plain_vanilla,
  Lee_Nagaosa_Wen, review}. Though the mechanism of the high Tc
superconductivity in these materials is still under debate, physicists have
reached some consensus, such that the physics is dominated by CuO$_2$ plane, and
the superconductivity gap has d-wave pairing symmetry and so on \cite{review}.
However, in a recent experiment, Zhong et. al. observed a nodeless U-shape gap with
STM on a monolayer CuO$_2$[CuO(1)] plane grown on a
Bi$_2$Sr$_2$CaCu$_2$O$_{8+\delta}$(BSCCO) substrates with MBE technique
\cite{xue}. They also found that the U-shape gap is robust against impurity
scattering and closed at a temperature near the Tc of the substrates. This
observation suggests a nodeless superconducting gap in the CuO(1) plane. However, it
is contradict with the well accepted d-wave pairing symmetry in cuprate which
has four nodes and V-shape gap in local density of states.

Soon after the experiment, Zhu et al. proposed that the almost same Tc between
CuO(1) layer and BSCCO suggests the superconducting gap observed in CuO(1) layer
is induced by proximity effect, while the gap is U-shape because of the
multi-band nature in the CuO(1) layer\cite{zhu}. They argued that the hole
transfer between the surface CuO(1) layer and bulk BSCCO is not significant, so
the substrate remains charge neutral, while the hole concentration in CuO(1) is
one hole per oxygen. Thus the hole concentration in CuO(1) is much larger than
the one in CuO$_2$ plane in the cuprate. Such a large hole doping makes CuO(1)
layer a good metal instead of a doped Mott insulator. So instead of the t-J
model based on Zhang-Rice singlet, Zhu et al. considered a phenomenological
two-band model of oxygen 2p$_x$ and 2p$_y$ orbitals with proximity induced
intraorbital pairings. The pairing was described by three phenomenological
parameters $\Delta_0$, $\Delta_x$, and $\Delta_y$ corresponding to on-site and
next nearest neighbor (NNN) pairing between two oxygen 2p$_x$ orbitals or two
2p$_y$ orbitals respectively. They studied the case where both $\Delta_0$ and
$\Delta_x + \Delta_y$ are positive, and their results showed that a nodeless gap
could be induced when $\Delta_0$ is large and inter-orbital hopping is small at
various electron density (0.6, 0.695, and 0.815 electron per oxygen).

In this paper, we perform a detailed investigation on the proximity effect
between the CuO(1) layer and the nearest CuO$_2$ layer in the BSCCO [CuO(2)]. We
estimate the pairing terms up to 4th nearest neighbor(NN) pairing with a
microscopic model. We find that the phenomenological parameter $\Delta_0$ and
$\Delta_x + \Delta_y$ should have opposite signs which are not studied in ref
\cite{zhu}. Using the pairing parameters with opposite signs, we obtain a
different phase diagram from Zhu et al. This result is a complementary to the
calculation of Zhu et al. We also investigate the effect of 3rd NN and 4th NN
terms. The phase diagram changes dramatically when 3rd NN term is included. When
the pairing is not very weak, whether the gap is nodal or nodeless is determined
by the chemical potential and is independent with the pairing strength. Our
results show that one can only observe U-shape gap when the electron
concentration is very low. Because the hole concentration in CuO(1) layer is
around 1 hole per oxygen, our results suggest a phase separation in the CuO(1)
layer, i.e. the hole concentration is very large in some region and very small
in others. This is consistent with the experimental observation, where the
U-shape gap is only observed in some regions while a pseudogap like behavior is
observed in other regions. We also check the effect of interorbital pairing
described by the 4th NN term, and the result shows that it is neglectable in
terms of the phase diagram.

The paper is organized as follows. In Sec. II, we present our detailed analysis
to proximity effect and introduce our model. In Sec. III, we discuss the
numerical results of the phase diagrams and analyze the resultant phase diagrams
based on analytical derivation. Our conclusions can be found in Sec. IV.

\section{MODEL}
\begin{figure}[htb]
  \includegraphics[width=0.5\textwidth]{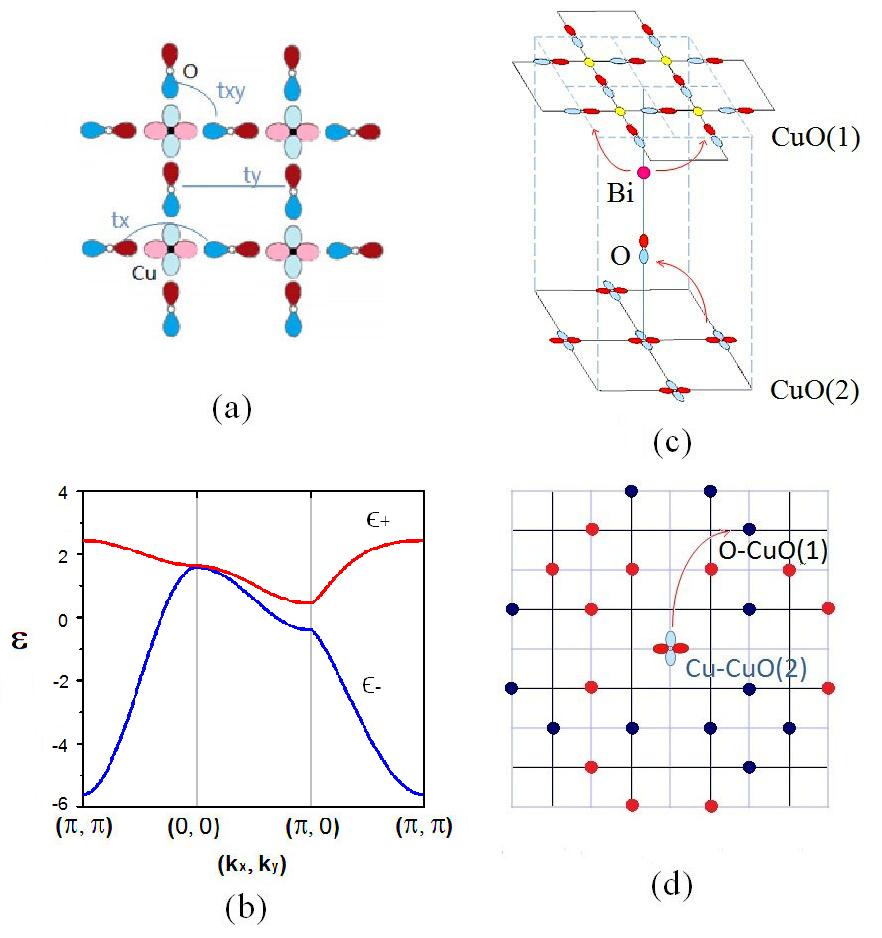}
  \caption{(a) The Schematic diagram of hopping integrals of the two band model.
    (b) The band structure with $t_{xy}= 1$, $t_x= 0.5$, and $t_y= 0.3$. (c) the
    tunneling path between CuO(1) and CuO(2) layers. (d) The relative sign of
    the tunneling matrix element, red represents positive sign while black
    represents negative sign. See text for details.}
  \label{fig:1}
\end{figure}

We start with a similar two-band model with the one used in ref. \cite{zhu},
\begin{align}
\label{eq:1}
H & = H_0 + H_p,
\end{align}
where $H_0$ describes the kinetic energy of CuO(1) layer and reads
\begin{align}
\label{eq:2}
H_0 & = \sum_{\alpha \beta \mathbf{k} \sigma} \epsilon_{\alpha \beta}(\mathbf{k}) c_{\mathbf{k} \alpha \sigma}^{\dag} c_{\mathbf{k} \beta \sigma}.
\end{align}
Here, $c_{\mathbf{k} \alpha \sigma}$ is the annihilation operator of electron
with wavenumber $\mathbf{k}$ orbital $\alpha$ and spin $\sigma$. $\alpha, \beta
= x, y$ are the orbital indices which correspond to oxygen $2p_x$ and $2p_y$
orbital respectively. Following Zhu et al. \cite{zhu}, we consider the simplest
case with NN and NNN hopping only,
\begin{align}
  \label{eq:3}
  \epsilon_{xx}(\mathbf{k}) & = 2 (t_x \cos k_x + t_y \cos k_y) - \mu \nonumber\\
  \epsilon_{yy}(\mathbf{k}) & = 2 (t_y \cos k_x + t_x \cos k_y) - \mu \nonumber\\
  \epsilon_{xy}(\mathbf{k}) & = \epsilon_{yx}(\mathbf{k}) = 4 t_{xy} \sin \frac{k_x}{2} \sin \frac{k_y}{2},
\end{align}
where $t_x$, $t_y$, and $t_{xy}$ are hopping integrals as shown in fig.
\ref{fig:1}(a), and $\mu$ is the chemical potential. Note that the different form
of $\epsilon_{\alpha\beta}$ between the above and the one used in Zhu et al. is
because of different choices of the orbital orientations.

Here, we consider a two-band model with oxygen 2p$_x$ and 2p$_y$ orbitals
instead of t-J model or Hubbard model because the hole concentration in CuO(1)
layer is much larger than the one in the usual CuO$_2$ layer of cuprate. The
hole concentration is around 1 hole per oxygen in the former one and usually no
more than 0.15 hole per oxygen in the latter one. It is more natural to start
with the oxygen 2p bands instead of Zhang-Rice band. In this scenario, the
electron configuration of Cu ion is still 3d$^9$ and the electron on the Cu site
behaves still like a localized spin. As pointed out by Zhu et al., the coupling
between the spin on Cu and oxygen bands may lead to renormalizations of the
hopping integrals as in Kondo lattice system \cite{zhu}. Therefore in the calculations
below, we treat them as phenomenological parameters instead of using the bare
values.

As discussed in the introduction, the same Tc of CuO(1) and BSCCO substrate
indicates that the superconductivity in CuO(1) is induced by proximity effect.
So the pairing term $H_p$ reads
\begin{small}
\begin{align}
  \label{eq:4}
  H_p & = \frac{1}{2}\sum_{i \alpha, j \beta} \Delta^{*}_{\alpha\beta}(\mathbf{r}_j - \mathbf{r}_i) (c_{i \alpha \uparrow} c_{j \beta \downarrow} - c_{i \alpha \downarrow}c_{j \beta \uparrow}) + h.c,
\end{align}
\end{small}
where $\Delta_{\alpha \beta}$ tracks the proximity effect due to the CuO(2)
layer. $\Delta_{\alpha \beta}$ should be proportional to the superconducting
order parameter in CuO(2) layer and the tunneling matrix element between the two
layers
\begin{align}
  \label{eq:5}
\Delta_{\alpha \beta}(\mathbf{R}) & \propto \sum_{i' j' \alpha \beta} T_{i \alpha,i'}T_{i + \mathbf{R} \beta,j'} \Delta^{(2)}_{i' j'},
\end{align}
where $T_{i \alpha, i'}$ is the matrix element of tunneling between Cu
3d$_{x^2-y^2}$ orbital at site $i'$ in CuO(2) plane and the oxygen $\alpha$
orbital at site $i$ in CuO(1) plane, $\Delta^{(2)}_{i' j'}$ is the
superconducting pairing of two holes at site $i'$ and $j'$ in CuO(2) layer. Because
of the d-wave pairing in the CuO(2) plane, one have $\Delta^{(2)}_{i' j'} =
\Delta^{(2)} (\delta_{j', i' \pm x} - \delta_{j', i' \pm y})$, where
$\Delta^{(2)}$ is the d-wave pairing order parameter in CuO(2) plane. In the
following, we choose the gauge that $\Delta^{(2)}$ is real.

Then we consider the tunneling matrix element $T_{i \alpha, i'}$. Because of the
large spatial distance between the two CuO$_2$ layers, direct hoppings are very
difficult. Thus the hopping of a hole from CuO(2) to CuO(1) is consists of three
steps $Cu \rightarrow O \rightarrow Bi \rightarrow O$ as shown in
fig.~\ref{fig:1}(c). In the first step, a hole on Cu 3d$_{x^2-y^2}$ orbital hops
to an apical O 2p$_z$ orbital in SrO plane. However, as pointed out by Yan Chen et
al. \cite{yan}, it is forbidden for a hole to hop to the 2p$_z$ orbital of the
oxygen just above it. A hole can only hop to nearest neighbor apical oxygen
2p$_z$ orbitals as shown in fig.~\ref{fig:1}(c). In the second step, the hole
hops from O 2p$_z$ orbital to 6s orbital of the Bi just above it. Finally,
it hops from Bi 6s orbital to O 2p$_x$/2p$_y$ orbital in the monolayer CuO$_2$
in the third step. Similar to the first step, because Bi is in the center of
each plaquette of the CuO(1), the direct hopping between Bi 6s orbital to
nearest neighbor O 2p$_x$/2p$_y$ orbitals is forbidden. Thus the hole hops from Bi
to the eight next nearest neighbor O orbitals as shown in fig.~\ref{fig:1}(c).

In summary, a hole at a Cu 3d$_{x^2-y^2}$ orbital in CuO(2) plane could hop to
24 different oxygen sites in CuO(1) plane as shown in fig.~\ref{fig:1}(d). It is
obviously that all the $T_{i \alpha, i'}$ for a given $i'$ have same amplitude,
but they may have different signs. Therefore we could define $T_{i \alpha, i'} =
s_{i \alpha, i'}T_0$, where $T_0$ is the amplitude while $s$ tracks the sign.
$T_0$ is hard to be calculated because of the lack of knowledge of the hopping
details, while $s$ could be calculated based on the analysis of the orientations
of the orbitals involved. In fig.~\ref{fig:1}(d), we depict $s$ for a given Cu
site $i'$, and the orientation of the orbitals are depicted in
fig.~\ref{fig:1}(a) and (c). Note that in the analysis above, we assume that
hole could only hop to 6s orbital of Bi. Because of the different sign
structures between the 6p$_z$ and 6s orbitals, their contributions to the
tunneling matrix element $T_{i \alpha, i'}$ have opposite signs. Therefore both
$T_0$ and $s$ could change if the Bi 6p$_z$ orbital is also involved in the
hopping process. However, the relative sign between $s_{i \alpha, i'}$ and $s_{j
  \beta, i'}$ does not change, and we will show it below that only the relative
sign is important.

Based on the analysis above, Eq.~(\ref{eq:5}) could be rewritten as
\begin{align}
\label{eq:6}
  \Delta_{\alpha \beta}(\mathbf{R}) \propto \sum_{i' j' \alpha \beta} s_{i \alpha,i'} s_{i+\mathbf{R} \beta,j'} (\delta_{j', i' \pm x} - \delta_{j', i' \pm y}),
\end{align}
where $s$ is the sign due to the hopping of holes, and the term in the
parenthesis tracks the d-wave pairing symmetry in the BSCCO substrate. It is
obviously that only the relative sign of different $s$ is important. Our results
show that a hole on one oxygen site could paired with another hole on up to
24th NN oxygen site. For simplicity, we consider only up to the 4th NN, and the
Fourier's transformation of $\Delta_{\alpha \beta}$ reads
\begin{align*}
  \Delta_{xx}(k_x, k_y) & =-\Delta_{yy}(k_y, k_x) = \Delta_{0}+\Delta_{2}(k_x, k_y) + \Delta_3(k_x, k_y) \\
  \Delta_{xy}(k_x, k_y) & = \Delta_{yx}(k_y, k_x)= \Delta_4(k_x, k_y),
\end{align*}
where $\Delta_{0} =24 \Delta$, $\Delta_{2}(k_x, k_y) =48 \Delta\cos k_x-32
\Delta\cos k_y$, $\Delta_3(k_x, k_y) = -44 \Delta \cos k_x \cos k_y$, and
$\Delta_4(k_x, k_y) = 20 \Delta (\sin \frac{3 k_x}{2}\sin \frac{k_y}{2} - \sin
\frac{3 k_y}{2}\sin \frac{k_x}{2})$ which correspond to on-site, NNN, 3rd NN,
and 4th NN pairing respectively. Note that the NN term vanishes in above
analysis. $\Delta$ is a parameter tracking the strength of the proximity effect.
In previous study, Zhu et al. considered the $\Delta_0$, $\Delta_2$ terms by introducing
three phenomenological parameters $\Delta_0$, $\Delta_x$ and $\Delta_y$. They
analyzed the phase diagram with only positive $\Delta_0$ and $\Delta_x +
\Delta_y$\cite{zhu}. By comparing with their definitions, we find that the three
parameters are $\Delta_0 = 24 \Delta$, $\Delta_x = -24 \Delta$ and $\Delta_y =
16 \Delta$ respectively. This leads to opposite sign between $\Delta_0$ and
$\Delta_x + \Delta_y$ which is not discussed in their paper.

\section{Results and Analysis}

By diagonalizing the Hamiltonian
\begin{footnotesize}
  \begin{align}
    \label{eq:7}
  H & = \sum_{\mathbf{k}} \left( c^{\dag}_{\mathbf{k} x \uparrow} c^{\dag}_{\mathbf{k} y
      \uparrow} c_{-\mathbf{k} x \downarrow} c_{- \mathbf{k} y
      \downarrow} \right)
      \left(
      \begin{matrix}
        \epsilon_{xx} & \epsilon_{xy} & \Delta_{xx} & \Delta_{xy} \\
        \epsilon_{xy} & \epsilon_{yy} & \Delta_{xy} & \Delta_{yy} \\
        \Delta_{xx} & \Delta_{xy} & - \epsilon_{xx} & - \epsilon_{xy} \\
        \Delta_{xy} & \Delta_{yy} & - \epsilon_{xy} & - \epsilon_{yy}
      \end{matrix}\right)
      \left(
      \begin{matrix}
        c_{\mathbf{k} x \uparrow}
        \\
        c_{\mathbf{k}
          y \uparrow} \\
        c^{\dag}_{-\mathbf{k} x
          \downarrow} \\
        c^{\dag}_{- \mathbf{k} y \downarrow}
      \end{matrix}
\right),
\end{align}
\end{footnotesize}
the quasiparticle energy reads
\begin{align*}
\pm E_{u}(\textbf{k}) &=\pm \sqrt{\frac{A(\textbf{k})+\sqrt{A(\textbf{k})^2-4B(\textbf{k})}}{2}},\\
\pm E_{l}(\textbf{k}) &=\pm \sqrt{\frac{A(\textbf{k})-\sqrt{A(\textbf{k})^2-4B(\textbf{k})}}{2}},
\end{align*}
where $ A(\mathbf{k})
=\epsilon_{xx}^2+\epsilon_{yy}^2+2\epsilon_{xy}^2+\Delta_{xx}^2+\Delta_{yy}^2+2\Delta_{xy}^2$,
and $ B(\mathbf{k})
=\left(\Delta_{xy}^2-\epsilon_{xy}^2+\epsilon_{xx}\epsilon_{yy}-\Delta_{xx}\Delta_{yy}\right)^2
+\left(2\epsilon_{xy}\Delta_{xy}-\epsilon_{xx}\Delta_{yy}-\epsilon_{yy}\Delta_{xx}\right)^2
$. The node of quasiparticle energy corresponds to $B(\mathbf{k})= 0$ which
requires
\begin{align}
\label{eq:8}& \Delta_{xy}^2-\epsilon_{xy}^2+\epsilon_{xx}\epsilon_{yy}-\Delta_{xx}\Delta_{yy} = 0 \\
\label{eq:9}& 2\epsilon_{xy}\Delta_{xy}-\epsilon_{xx}\Delta_{yy}-\epsilon_{yy}\Delta_{xx} = 0.
\end{align}
The phase diagram can be calculated by solving these two equations. For a given
set of parameters, the quasiparticle spectrum has node if one can find at least
one solution of above equations, and it is nodeless if there is no solution for
the above equations.

\begin{figure}[htbp]
  \includegraphics[width=0.5\textwidth]{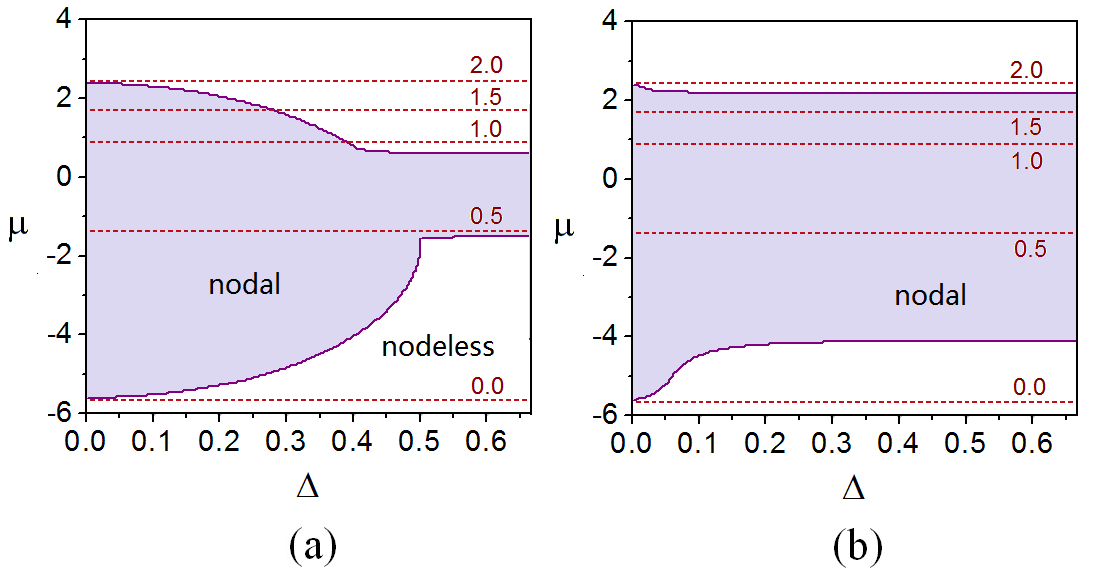}
  \caption{Phase diagram for nodal and nodeless superconducting gap with
    $t_{xy}= 1, t_x= 0.5, t_y= 0.3.$. Dashed lines: the chemical potentials
    corresponds to 0, 0.5, 1, 1.5, and 2 electrons per oxygen respectively. (a)
    on-site and NNN terms only. (b) 3rd NN term is included. }
  \label{fig:2}
\end{figure}

At first, we consider the case with only on-site and NN terms where the
interorbital pairing $\Delta_{xy} = 0$. This is similar to the case studied in
ref. \cite{zhu}, except that $\Delta_0$ and $\Delta_x + \Delta_y$ have different
signs now. To compare with the result of Zhu et al., we choose same set of
parameters of hopping integrals, i.e. $t_{xy} = 1$, $t_x = 0.5$, and $t_y =
0.3$. The corresponding band structure is presented in fig.~\ref{fig:1}(b), where
$\epsilon_{\pm}=\frac{\epsilon_x+\epsilon_y}{2}\pm\frac{\sqrt{(\epsilon_x-\epsilon_y)^2+4\epsilon_{xy}^2}}{2}$
 is the dispersion of the two bands respectively. The resultant
phase diagram is given in fig.~\ref{fig:2}(a). One could find that the region
with a nodeless gap is rather small when $\Delta$ is small and enlarges with
the increase of $\Delta$. However, even when the pairing term is rather large (note that
on-site pairing $\Delta_0 = 24 \Delta$), the gap is always V-shape for $-1.5 <
\mu < 0.75$. This is different from Zhu et al.'s result where a nodeless gap
could be observed when $\Delta_0$ is large for $\mu = 0.2, -0.4, -0.8$.

\begin{figure}[htbp]
  \includegraphics[scale=0.3]{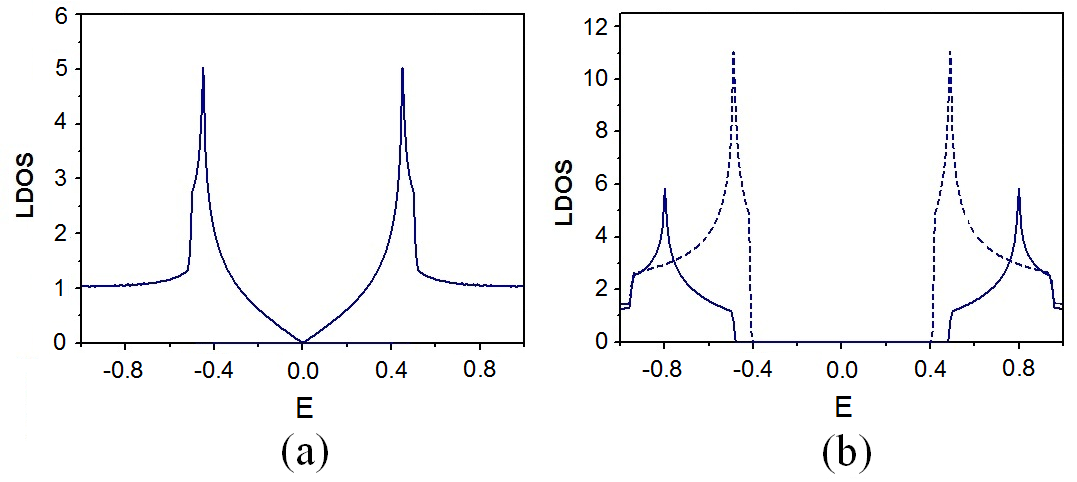}
  \caption{The quasi-particle local density of states at $\Delta =0.1, \mu =-5$.
    (a) on-site and NNN terms only, where the superconducting gap has nodes. (b)
    solid line: 3rd NN term is included; dashed line: both 3rd NN and 4th NN
    term are included.  Both cases have nodeless gap.}
  \label{fig:3}
\end{figure}

The phase diagram changes dramatically when the 3rd NN term $\Delta_3$ is
included as shown in fig.~\ref{fig:2}(b). In small $\Delta$ case, the nodeless
region is slightly enlarged. For example, as shown in fig.~\ref{fig:3} (a), with
only $\Delta_0$ and $\Delta_2$, the local density of states at $\Delta = 0.1$,
$\mu = -5$ is V-shape, which indicates the existence of gap nodes. However, it
becomes U-shape when $\Delta_3$ is included, as shown in fig.~\ref{fig:3}(b).
Meanwhile, the nodeless region is strongly suppressed by $\Delta_3$ in large
$\Delta$ case. According to fig.~\ref{fig:2}(b), when $\Delta > 0.2$, the phase
boundary is almost independent on $\Delta$, and a nodeless gap could be observed
if and only if $\mu > 2.24$ or $\mu < -4.08$ corresponding to less than
$0.12$ hole per oxygen or less than $0.15$ electron per oxygen.

\begin{figure}[htbp]
  \includegraphics[scale=0.16]{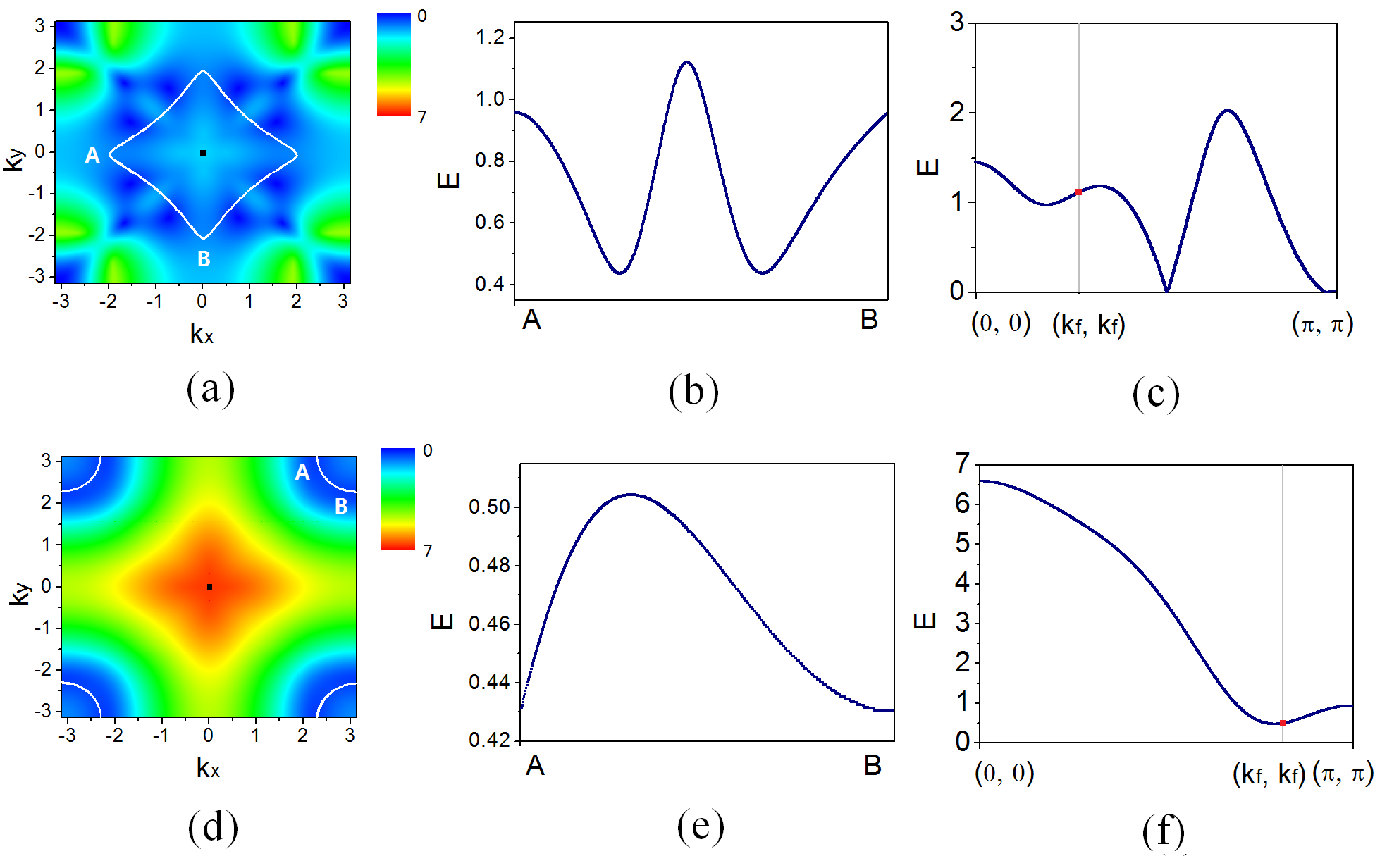}
  \caption{ Left panel: lower branch of quasiparticle spectrum $E_{l}(\mathbf{k})$
    in Brillouin zone. The white line is
    the Fermi surface. Middle panel: $E_{l}(\mathbf{k})$ on Fermi surface from
    point A to point B in (a) and (d) respectively. Right panel: $E_{l}(\mathbf{k})$
     along diagonal direction. The parameters are $t_x = 0.5$, $t_y = 0.3$,
    $t_{xy} = 1$, $\Delta = 0.1$, $\mu=0.2$(upper panel) and $\mu=-5$ (lower
    panel).}
  \label{fig:4}
\end{figure}

To further investigate the property of the superconducting gap, we depict the
lower branch of the quasiparticle spectrums $E_{l}(\mathbf{k})$
in whole Brillouin zone with $\Delta = 0.1$ at $\mu = 0.2$
(nodal gap region) and $\mu = -5$ (nodeless gap region) in fig.~\ref{fig:4}(a)
and (d) respectively. The white line is the underlining Fermi surface. The
$E_{l}(\mathbf{k})$ along Fermi surface are depicted in fig.~\ref{fig:4}(b) and
(e) respectively. One interesting phenomenon is that there is no node on
Fermi surface in $\mu = 0.2$ case. Instead, the nodes of the quasiparticle dispersion
are at points away from Fermi surface in the diagonal direction as shown in
fig.~\ref{fig:4}(c). This is different from usual case where the nodes are
always on the Fermi surface.

To understand these phenomena, we try to solve Eq.~\eqref{eq:8} and \eqref{eq:9}
analytically. We consider a special solution with $\cos k_x = \cos k_y$, i.e.
$\mathbf{k}$ is along diagonal direction, then one has $\epsilon_{xx} =
\epsilon_{yy}$ and $\Delta_{xx} = - \Delta_{yy} = (24 + 16 \cos k_x - 44 \cos^2
k_x) \Delta$. Therefore Eq.~\eqref{eq:9} can be automatically satisfied in the case
interorbital pairing $\Delta_{xy}$ vanishes. Then the Hamiltonian \eqref{eq:7}
in band basis reads
\begin{small}
  \begin{align*}
  H & = \sum_{\mathbf{k}} \left( c^{\dag}_{\mathbf{k} (-) \uparrow} c^{\dag}_{\mathbf{k} (+)
      \uparrow} c_{-\mathbf{k} (-) \downarrow} c_{- \mathbf{k} (+)
      \downarrow} \right) H(\mathbf{k})
      \left(
      \begin{matrix}
        c_{\mathbf{k} (-) \uparrow}
        \\
        c_{\mathbf{k} (+) \uparrow} \\
        c^{\dag}_{-\mathbf{k} (-) \downarrow} \\
        c^{\dag}_{- \mathbf{k} (+) \downarrow}
      \end{matrix}
\right),
\end{align*}
\end{small}
with
\begin{align}
\label{eq:10}
  H(\mathbf{k}) & = \left(
      \begin{matrix}
        \epsilon_{-} - \mu & 0 & 0 & -\Delta_{xx}\\
        0 & \epsilon_{+} - \mu & -\Delta_{xx} & 0 \\
        0 & -\Delta_{xx} & - \epsilon_{-} + \mu & 0 \\
        -\Delta_{xx} & 0& 0 & - \epsilon_{+} + \mu
      \end{matrix}\right),
\end{align}
where $\epsilon_{\pm}(\mathbf{k}) = 2(t_x + t_y)\cos k_x \pm |4 t_{xy} \sin k_x
/ 2 \sin k_y / 2|$ for $\cos k_x = \cos k_y$. Then the quasiparticle energy
reads
\begin{align*}
  \pm E_{u}(\textbf{k})&= \pm \left\{ \delta {\epsilon}(\mathbf{k})  +
  \sqrt{\left[\bar{\epsilon}(\mathbf{k})-\mu\right]^2+\Delta^2_{xx}(\mathbf{k})}\right\}.\\
  \pm E_{l}(\textbf{k})&= \pm \left\{ \delta {\epsilon}(\mathbf{k})  -
  \sqrt{\left[\bar{\epsilon}(\mathbf{k})-\mu\right]^2+\Delta^2_{xx}(\mathbf{k})}\right\}.\\
\end{align*}
where $\delta \epsilon = (\epsilon_+ - \epsilon_-)/2$ and $\bar{\epsilon} =
(\epsilon_+ + \epsilon_-) / 2$. Note that Eq.~\eqref{eq:10} indicates that the
intraband pairing vanishes while the interband pairing dominants at diagonal
direction. The interband pairing leads to the additional term $\delta \epsilon$
in the quasi-particle energy which is responsible for the shift of gap minimum
away from Fermi surface.

\begin{figure}[htbp]
  \includegraphics[scale=0.25]{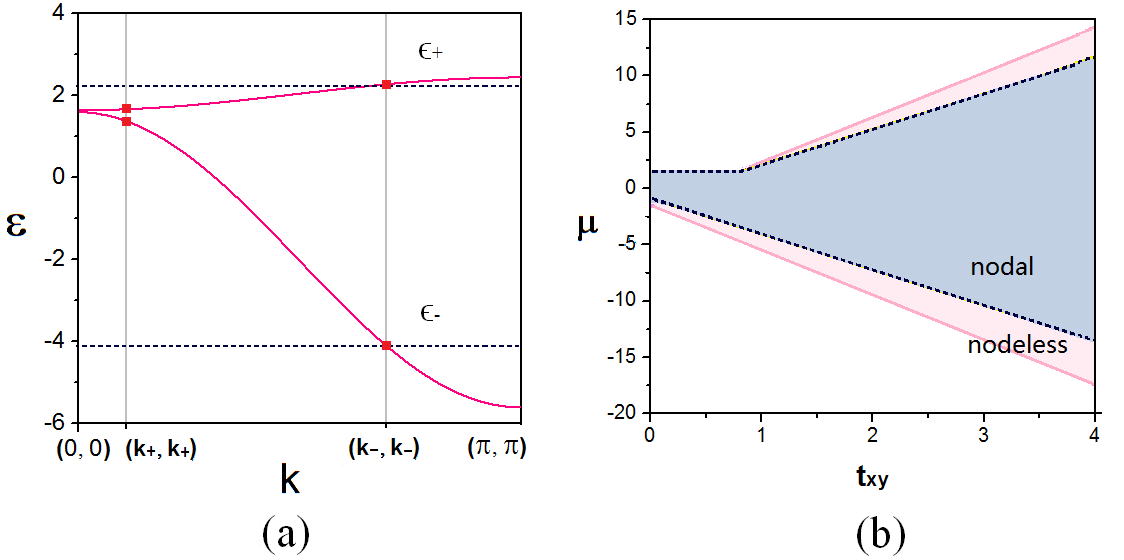}
  \caption{(a) $k_{\pm}$ and $\epsilon_{\pm}(k_{\pm}, k_{\pm})$ for
    $t_{xy}=1,t_x=0.5, t_y = 0.3$. The dashed line is the phase boundary between
    nodal and nodeless region. (b) the phase diagram at $\Delta = 1$ and $t_x +
    t_y = 0.8$. The dashed line is given by $\max[ \epsilon_{+}(k_+, k_+),
    \epsilon_{+}(k_-, k_-)]$ and $\min[\epsilon_{-}(k_+, k_+), \epsilon_{-}(k_-,
    k_-)]$. See text for details.}
  \label{fig:5}
\end{figure}

Then we investigate the $\Delta$ independence of the phase boundary. This could
be understood by considering the quasiparticle energy at three special k-points,
$(\pi, \pi)$ and $\mathbf{k}_{\pm}$ where $\cos k_{x \pm} = \cos k_{y \pm}=
\frac{2 \pm \sqrt{70}}{11}$, and $\Delta_{xx}(\mathbf{k}_{\pm}) = 0$. It is
obvious that $E_{l}(\mathbf{k}_{\pm}) = \epsilon_+(\mathbf{k}_{\pm}) - \mu$ if
$\mu \ge \bar{\epsilon}$, and $E_{l}(\mathbf{k}_{\pm}) = \mu -
\epsilon_-(\mathbf{k}_{\pm})$ if $\mu < \bar{\epsilon}$. Therefore at least one of
$E_{l}(\mathbf{k}_{\pm})$ is positive, if $\mu$ is between
$\epsilon_{\pm}(\mathbf{k}_+)$ or $\epsilon_{\pm}(\mathbf{k}_-)$. On the other
hand, $E(\pi, \pi) < 0$ if $\Delta$ is large enough. This means that $E_{l}$ has
at least one node if $\epsilon_-(\mathbf{k}_+) \le \mu \le
\epsilon_+(\mathbf{k}_+)$ or $\epsilon_-(\mathbf{k}_-) \le \mu \le
\epsilon_+(\mathbf{k}_-)$ when $\Delta$ is large enough. In fig.~\ref{fig:5}(a),
we depict the phase boundary for chemical potential at $\Delta = 1$, $t_{xy} =
1$, $t_x = 0.5$, and $t_y = 0.3$. It is consistent with our analysis very well.

Since the phase boundary of the nodeless region depends only on the
$\epsilon_{\pm}(\mathbf{k}_{\pm})$ in diagonal direction, it depends only on the
ratio of $t_{xy} / (t_x + t_y)$. Thus we perform similar calculations for various
$t_{xy}$ with $t_x + t_y = 0.8$ and $\Delta = 1$ to check the effect of kinetic
energy. The results are presented in fig.~\ref{fig:5}(b) where the SC gap is
nodal if $\mu$ lies in the blue region, and nodeless if $\mu$ lies in the red
region. The dashed lines are given by $\max[ \epsilon_{+}(k_+, k_+),
\epsilon_{+}(k_-, k_-)]$ and $\min[\epsilon_{-}(k_+, k_+), \epsilon_{-}(k_-,
k_-)]$. They coincide very well with the phase boundary from numerical
calculations. According to the figure, the gap could be nodeless only when $\mu$
is close to top of the $\epsilon_+$ band or the bottom of the $\epsilon_-$ band.

Finally, we also include the 4th NN term $\Delta_4$ to study the effect of
inter-orbital pairing. A typical result is shown as dashed line in
fig.~\ref{fig:3}(b). Comparing to the result without 4th NN term (the solid
line), the lineshape of the local density of states is different, and the gap is
slightly suppressed, but the resultant phase diagram is almost the same as the one
without $\Delta_4$. Therefore in terms of the phase diagram, the interorbital term
$\Delta_{xy}$ can be safely ignored.

\section{Summary and Discussions}

We now discuss the possible relation between our results and the experimental
observation of the nodeless gap in ref. \cite{xue}. In our results, the nodeless
gap can exist when the proximity pairing strength $\Delta_0$ is comparable to
the hopping integrals. This is possible because of the renormalization of the
oxygen band by coupling to localized spin on Cu, as discussed by Zhu et
al.\cite{zhu}. Our results also show that a nodeless gap could only exist at
very large or very small hole concentrations. However, in the low hole concentration
regime,  the holes on oxygen will form Zhang-Rice singlets with the spins on Cu
\cite{Zhang_rice}, and the holes can be effectively considered as doped on Cu
sites.
Therefore our model is not valid in this case, which means that the low hole
concentration regime should be excluded from consideration. Thus one can only have
a nodeless gap when the hole concentrations is very large. This can not be
satisfied if the monolayer CuO$_2$ is homogeneous because there is only 1 hole
per oxygen. However, the experimental data shows that there are actually two kinds of
regions, one has a large pseudogap-like V-shape gap, and the other has a
superconducting U-shape gap \cite{xue}. Therefore there may be a phase separation in
the system, where one kind of region with low hole concentration exhibits
psuedogap behavior, and the other kind with very large hole concentration exhibits
nodeless superconducting gap.

In the calculations, a few assumptions have been introduced.  For example, we have assumed that the
charge transfer between the surface monolayer CuO$_{2}$ and the substrate is not significant.
Therefore the average hole concentration in the CuO(1) is close to 1 hole per
oxygen. We also assumed that the main effect of localized
spins on Cu sites is to renormalize the oxygen bands through
a Kondo lattice like physics, so we can consider an effective
phenomenological model with only oxygen bands.  Our analysis of the nodeless gap depends on these assumptions.
Though these assumptions are difficult to check theoretically, they can be tested experimentally.

In summary, based on a detailed analysis of the hopping process for a hole
between surface CuO$_2$ plane and an inner CuO$_2$ plane, we estimate the signs
of the pairing parameters in the CuO$_2$ plane by using a phenomenological
proximity Hamiltonian. Our calculation complete the proximity-induced-pairing
scenario and show that nodeless gap could be induced only when the hole
concentration on the monolayer CuO$_2$ is very large. This can give a further
experimental test towards the proximity scenario. We argue that the nodeless gap
could be related to the one observed in the experiment if there is phase
separation in the monolayer CuO$_2$.

\begin{acknowledgments}
  We would like to thank F. C. Zhang for very helpful discussions. This work was
  supported by NSFC 11674151, and The National Key Research and Development
  Program of China (No. 2016YFA0300300).
\end{acknowledgments}

\end{document}